\documentclass[twocolumn]{aa}
\usepackage{graphicx}
\usepackage{txfonts}
\begin{document}
   \title{Deep near-infrared imaging of RDCS J1252.9-2927 at z=1.237}

   \subtitle{The colour-magnitude diagram \thanks{Based on
   observations obtained at the European Southern Observatory using
   the ESO Very Large Telescope on Cerro Paranal (ESO program
   166.A-0701(B)) and the ESO New Technology Telescope on Cerro La
   Silla (ESO program 61.A-0676(A)).}}

   \author{C. Lidman\inst{1}
          \and
          P. Rosati \inst{2}
          \and
          R. Demarco \inst{2,3}
          \and
          M. Nonino \inst{4}
          \and
          V. Mainieri \inst{2}
	  \and
	  S. A. Stanford \inst{5,6}
          \and
          S. Toft \inst{7}
          }

   \offprints{C. Lidman: clidman@eso.org}

   \institute{European Southern Observatory, Alonso de Cordova 3107,
              Casilla 19001, Santiago, Chile
         \and
             European Southern Observatory, Karl-Scwarzschild-Strasse 2, 
              D-85748 Garching, Germany
         \and
             Institut d'Astrophysique de Paris, 98bis Boulevard Arago, 
              F-75014 Paris, France
	 \and 
             Instituto Nazionale di Astrofisica, Osservatorio Astronomico 
             di Trieste, via G.B. Tiepolo 11, I-34131, Trieste, Italy
         \and
             Department of Physics, University of California at Davis, 
             1 Shields Avenue, Davis, CA 95616, USA 
	 \and
             Institute of Geophysics and Planetary Physics, Lawrence 
             Livermore National Laboratory, L-413, Livermore, CA 94551, USA
         \and
             Astronomical Observatory, University of Copenhagen,
             Juliane Maries Vej 30, DK-2100 Copenhagen, Denmark}

   \date{Received 8 August 2003; accepted 14 October 2003}

   \abstract{ We present deep SofI and ISAAC near-infrared imaging
data of the X-ray luminous galaxy cluster \object{RDCS
J1252.9-2927}. The ISAAC data were taken at the ESO Very Large
Telescope under very good seeing conditions and reach limiting Vega
magnitudes of 25.6 and 24.1 in the $J-$ and $K_{\mathrm s}-$bands
respectively. The image quality is 0\farcs45 in both passbands. We use
these data to construct a colour-magnitude (C-M) diagram of galaxies
that are within 20\arcsec\ of the cluster center and brighter
than $K_{\mathrm s}=24$, which is five magnitudes fainter than
the apparent magnitude of a $L^{\star}$ galaxy in this cluster. The
C-M relation is clearly identified as an over-density of galaxies with
colours near $J-K_{\mathrm s}=1.85$. The slope of the relation is
$-0.05 \pm 0.02$ and the intrinsic scatter is 0.06 magnitudes with a
90\% confidence interval that extends from 0.04 to 0.09 magnitudes.
Both the slope and the scatter are consistent with the values measured
for clusters at lower redshifts. These quantities have not evolved
from $z=0$ to $z=1.24$.  However, significant evolution in the mean
$J-K_{\mathrm s}$ colour is detected. On average, the galaxies in
\object{RDCS J1252.9-2927} are 0.25 magnitudes bluer than early-type
galaxies in the Coma cluster. Using instantaneous single-burst
solar-metallicity models, the average age of galaxies in the center of
\object{RDCS J1252.9-2927} is 2.7 Gyrs.

    }

   \titlerunning{Deep near-infrared imaging of RDCS J1252.9-2927}

   \maketitle
%

\section{Introduction}

The tight relation between colour and apparent magnitude for
early-type galaxies in massive galaxy clusters (the C-M relation) is
seen at all redshifts, from the nearest clusters (Bower, Lucy \& Ellis
\cite{Bower92}) to the most distant clusters currently known (Rosati
et al. \cite{Rosati99}, Nakata et al. \cite{Nakata01}; van Dokkum et
al. \cite{vanDokkum01b}; Stanford et al. \cite{Stanford02}; Blakeslee
et al. \cite{Blakeslee03}, hereafter BFP).

Observations show that although the zero-point of the C-M relation
evolves considerably with increasing redshift (Arag\'{o}n-Salamanca et
al. \cite{Aragon93}; Stanford et al. \cite{Stanford98}, hereafter
SED; van Dokkum et al. \cite{vanDokkum01b}; Stanford et
al. \cite{Stanford02}; BFP), the slope of the relation and the
scatter about it appear to evolve very little (Ellis et
al. \cite{Ellis97}; SED, van Dokkum et al. \cite{vanDokkum00};
BFP). However, for some clusters at $z \sim 1$, there is tentative
evidence for a flattening in the slope (van Dokkum et
al. \cite{vanDokkum01b}; Stanford et al. \cite{Stanford02}).

Complimentary studies of clusters up to $z \sim 1 $ show that
significant evolution is also occurring in the galaxy luminosity
function (De Propris et al. \cite{DePropris98}; Massarotti et
al. \cite{Massarotti03}; Toft, Soucail \& Hjorth \cite{Toft03b}) and
the fundamental plane (van Dokkum et al. \cite{vanDokkum98}, van
Dokkum et al. \cite{vanDokkum03}).  Early-type galaxies in rich galaxy
clusters are uniformly becoming both brighter and bluer as they become
younger.

The monolithic collapse scenario of Eggen, Lynden-Bell \& Sandage
(\cite{Eggen62}) is an attractive framework to model the
observations. In this scenario, the bulk of the stars form in a single
burst over a relatively short period of time at redshifts greater than
two (Ellis et al. \cite{Ellis97}; Bower, Lucy and Ellis,
\cite{Bower92}). Hence, the scatter about the C-M relation and the
slow steady evolution in the colours and luminosities of early-type
galaxies are explained by the great age of the bulk of the
stars. Extensions to this model allow future merging and star
formation, but the scatter in the C-M relation limits the amount of
merging and star formation that can take place (Bower, Kodama \&
Terlevich \cite{Bower98}).

This picture of passively-evolving, very old early-type galaxies in
massive clusters should be compared to observations of moderately
distant clusters ($z=0.2$ to $z=0.8$) which show that the fraction of
late-type, star-forming galaxies in massive clusters increases with
redshift (Dressler et al. \cite{Dressler97}; Couch et
al. \cite{Couch98}, van Dokkum et al. \cite{vanDokkum00}; Nakata et
al. \cite{Nakata01}), while the fraction of early-type galaxies does
the opposite (Treu et al. \cite{Treu03b}; van Dokkum et
al. \cite{vanDokkum00}). This picture should also be compared to
semi-analytic and direct {\sl N}-body numerical simulations which show
that galaxy formation and evolution involves ubiquitous merging at all
epochs (Kauffman and Charlot, \cite{Kauffmann98}; Cole et
al. \cite{Cole00}; Pearce et al. \cite{Pearce01}). The simulations are
able to reproduce the slope in the C-M relation and the scatter about
it in present day clusters, although some difficulties, particularly
at the high mass end, remain (Cole et al. \cite{Cole00}). At higher
redshifts, the slope is predicted to flatten and the scatter is
predicted to stay approximately constant, although there is a slight
increase in the scatter at the bright-end of the C-M relation for
clusters with $z>1$ (Kauffman \& Charlot, \cite{Kauffmann98}; Ferreras
\& Silk \cite{Ferreras00}).

Thus, there are two quite distinct pictures for the formation of
early-type galaxies. In hierarchical merger models, the bulk of the
stars form in disk-like galaxies that later merge to become
early-type galaxies. In monolithic collapse models, the bulk of the
stars form in early-type galaxies and subsequent merging and star
formation are limited. In both models, the C-M relation is
fundamentally a relation between the mass of a galaxy and
the average metallicity of the stellar population (Faber
\cite{Faber73}).  In hierarchical merger models,
large ellipticals are formed from large spirals, which are better able
to retain the metals that result from stellar evolution (Kauffman and
Charlot \cite{Kauffmann98}). Similarly, in monolithic collapse models,
larger ellipticals are better able to retain their metals. Although
age differences can be used to explain the slope of the C-M relation
at low redshifts, the slope and the C-M relation itself are lost by
$z=0.2$ (Kodama \& Arimoto \cite{Kodama97}) if age is the sole reason
for the slope.

The morphological evolution that is seen in hierarchical models can
lead to a bias (the progenitor bias) in morphologically selected
samples (van Dokkum et al. \cite{vanDokkum00}; van Dokkum et
al. \cite{vanDokkum01a}).  The bias causes the progenitors of the
youngest low-redshift ellipticals to drop out of morphologically
selected high-redshift samples. Consequently, the C-M relation is
similar to that of a single-age stellar population formed at very high
redshift and the scatter in the relation is approximately redshift
independent. The progenitor bias is implicitly included in the
semi-analytical simulations described above and allows the star
formation history of early-type galaxies in clusters to be
considerably more varied than that in monolithic collapse models. The
model predicts that the fraction of early type galaxies in clusters
decreases with increasing redshift.

The importance of progenitor bias depends on the origin of the scatter
in the C-M relation.  If the scatter is entirely caused by age
differences, then progenitor bias is important. If the scatter is
partially caused by other effects, such as dissipationless merging
with little subsequent star formation (van Dokkum \& Ellis,
\cite{vanDokkum03b}) or metallicity, then progenitor bias becomes less
important and both the average age of the galaxies and the degree to
which galaxies form coevally increase. The importance of progenitor
bias also depends on the method used to derive C-M relations. C-M
relations that are derived from morphological catalogues are more
likely to be biased than C-M relations that are derived from
photometric or complete spectroscopic catalogues.

Although hierarchical models have become the standard model for
describing the formation of early-type galaxies in both cluster and
field environments, these models are unable to describe all the
observational data. Whereas the hierarchical merging model predicts a
dramatic difference in the star formation histories of early-type
galaxies in the field and in clusters (Diaferio et
al. \cite{Diaferio01}), only small differences are inferred from
observational data (Willis et al. \cite{Willis02}; Treu et
al. \cite{Treu02}; Treu \cite{Treu03a}; van Dokkum \& Ellis
\cite{vanDokkum03b}).  More stringent tests of hierarchical models
will come from observations of field and cluster galaxies beyond $z
\sim 1$.

In this paper we describe deep near-infrared (NIR) observations of
\object{RDCS J1252.9-2927}, an X-ray luminous cluster of galaxies at
z=1.237 (Rosati et al. \cite{Rosati03a}). These observations allow us
to construct a NIR C-M diagram of one of the most distant massive
clusters known to an unprecedented depth and accuracy.  Throughout
this paper, we assume $\Omega_{\mathrm M}=0.3$, $\Omega_{\mathrm
\Lambda}=0.7$ and $H_{\mathrm 0} = 70$ km/s/Mpc. In this cosmology,
1\arcmin\ on the sky corresponds to approximately 0.5 Mpc at
$z=1.237$. Unless specified otherwise, all colours and magnitudes are
on the 2MASS system.

\section{Observations}

\subsection{SofI NIR imaging}

As part of an NIR program to confirm high redshift galaxy clusters in
the ROSAT Distant Cluster Survey (Rosati et al. \cite{Rosati98}),
\object{RDCS J1252.9-2927} was observed during the nights of 1998 July
9th, 10th and 11th with SofI (Moorwood, Cuby \& Lidman,
\cite{Moorwood98}) on the ESO-NTT at the Cerro La Silla
Observatory. SofI is equipped with a Hawaii 1024x1024 HgCdTe array,
which, in the large field imaging mode, results in a pixel scale of
0\farcs29 per pixel and a field of view of 4\farcm9.  The observations
were done in $J$ and $K_{\mathrm s}$, which, at the redshift of the
cluster, approximately correspond to the rest frame $V-$ and
$z-$bands.
 
Individual exposures lasted 10 seconds in $K_{\mathrm s}$ and 20
seconds in $J$ and six of these were averaged to form a single
image. Between images, the telescope was offset by 10\arcsec\ to
30\arcsec\ in a semi-random manner.
 
\begin{figure}
\centering \includegraphics[width=8cm]{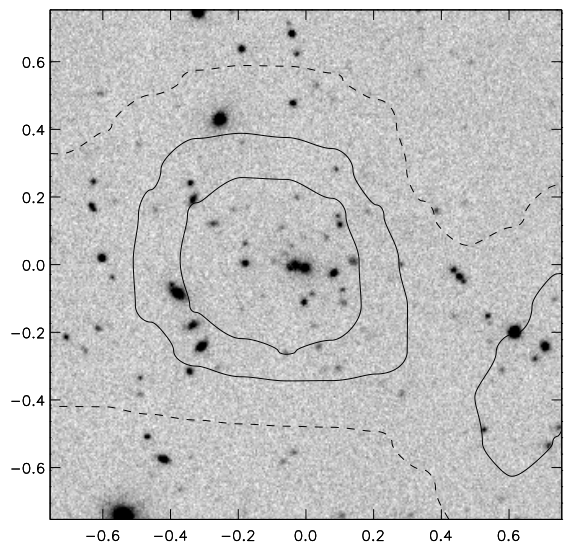}
\caption{The central part of the SofI $K_{\mathrm s}$-band image. The
3, 5 and 7$\sigma$ ROSAT X-ray contours are overlaid (Rosati et
al. \cite{Rosati98}). North is up and East is to the left. The image
is 90\arcsec\ on a side, which, for the adopted cosmology, is 0.75 Mpc
at $z=1.237$.}
\label{fig:image_SOFI}
\end{figure}

The data were reduced in the standard way. From each image, the
zero-level offset was removed, a flatfield correction was applied, and
an estimate of the sky from other images in the sequence was
subtracted.  Images with the best image quality, defined here as the
Full Width at Half Maximum (FWHM) of stellar sources, were then
registered and combined. A summary of the data is given in Table
\ref{tab:SOFI_observations}. The central part of the SofI $K_{\mathrm
s}$ band image is shown in Fig. \ref{fig:image_SOFI}.
 
The atmospheric conditions at the time the data were taken were very
good. Zero Points (ZP) were derived by observing standards from the
photometric catalogue of Persson et al.  (\cite{Persson98}). Several
standards were observed during each night. The scatter in the $J$ and
$K_{\mathrm s}$ ZPs throughout the entire run were less than 0.02
magnitudes. The SofI $J$ and $K_{\mathrm s}$ filters are a good match
to those used in the LCO (Persson) system, so no colour corrections
between the natural SofI system and the LCO (Persson) system are
made. However, we do transform the magnitudes and colours to the 2MASS
system (Carpenter \cite{Carpenter01})\footnote{See also
http://www.astro.caltech.edu/2mass.}. Since the 2MASS and LCO systems
are very similar, the transformations are small.

\begin{table}
\caption[SOFI data]{A summary of the observations taken with SofI.
The detection limit is the $5 \sigma$ detection threshold over an
1\farcs4 diameter aperture, which is approximately twice the stellar
FWHM.}
\label{tab:SOFI_observations}
\begin{tabular}{lrcc}
\hline\hline
Filter & Exposure & Image Quality  & Detection Limit\\
       & (seconds)     & (\arcsec )     & (Vega magnitudes) \\
\hline
$K_{\mathrm s}$     & 5400          & 0.68  & 21.1\\
$J$       & 5040          & 0.72  & 22.5\\
\hline 
\end{tabular}
\end{table}  

We used the SExtractor software (Bertin \& Arnouts \cite{Bertin96}) to
detect objects, to do the photometry and to classify sources as either
point like or extended. The colours are derived from the flux within
fixed apertures of 6 pixels (1\farcs73). A small correction (0.02
magnitudes) is applied to the J band data to account for the slightly
poorer image quality.  The total magnitude was estimated using the
``BEST'' magnitude in SExtractor (Bertin \& Arnouts
\cite{Bertin96}). Since many of the sources in the center of the
cluster are blended, the total magnitude was more often than not the
corrected isophotal magnitude.

The flux of non-stellar sources are corrected for galactic
extinction. Using $E(B-V)=0.075$ (Bouwens et al. \cite{Bouwens03}),
the corrections for $J$ and $K_{\mathrm s}$ are 0.067 and 0.027
magnitudes respectively. Stellar sources are not corrected for
galactic extinction.

About a dozen sources from the 2MASS point source catalogue are
visible in the SofI images. The photometry for objects brighter than
$J\sim12.5$ and $K\sim12$ in the SofI images is generally poor because
of detector non-linearity. However, for fainter objects, we can make a
comparison between SofI magnitudes and those in the 2MASS
catalogue. The variance weighted differences between the magnitudes in
the 2MASS catalogue and the SofI derived magnitudes for 9 objects in
common is $\Delta J = -0.01$ and $\Delta K_{\mathrm s} =0.02$.

\begin{figure}
\centering
\includegraphics[width=8cm]{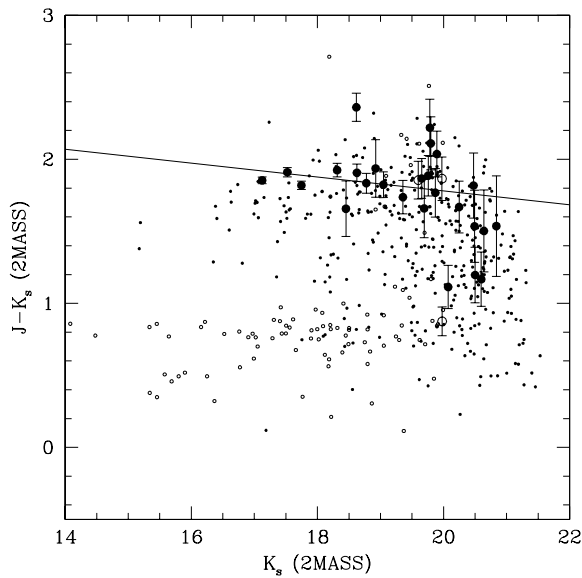}
\caption{The C-M diagram of objects in the SofI data. The large
symbols are objects that are within 20\arcsec\ of the cluster center
and the small symbols are objects that are beyond 60\arcsec\ of the
cluster center.  Objects that have been classified as stellar are
plotted with open symbols and non-stellar objects are plotted with
solid symbols. No classification is made for objects fainter than
$K_{\mathrm s}=20$. The solid line is a fit to the C-M relation that was
determined from the ISAAC data.}
\label{fig:CM_SOFI}
\end{figure}

The C-M diagram of objects within 20\arcsec\ of the cluster center is
shown in figure \ref{fig:CM_SOFI}. An over-density of galaxies with
$J-K_{\mathrm s} \sim 1.85$ is clearly seen in this diagram. Over the
entire SofI image other populations can be identified.  Stars have
$J-K_{\mathrm s}$ colours that vary between 0 and 1.0, with a well
defined peak at $J-K_{\mathrm s}=0.8$, which corresponds to early
M-dwarfs, and a less well defined peak at $J-K_{\mathrm s} \sim 0.5$,
which corresponds to mid-G-dwarfs.

The co-incidence of X-ray emission with two relatively bright galaxies
($K_{\mathrm s} \sim 17.5$) and the distinctive sequence of galaxies
with $J-K_{\mathrm s} \sim 1.85$ showed that \object{RDCS
J1252.9-2927} was probably a distant, rich galaxy cluster worthy of a
more detailed study. We therefore initiated a program to obtain deep
NIR images with ISAAC (this paper), deep optical and comprehensive
spectroscopic observations (Rosati et al. in preparation; Demarco
et al. in preparation) and deep X-ray observations (Rosati et
al. \cite{Rosati03a}).

\subsection{ISAAC NIR imaging}

\object{RDCS~J1252.9-2927} was re-observed with ISAAC (Moorwood et
al. \cite{Moorwood99}) on Antu (VLT-UT1) at the Cerro Paranal
Observatory. The observations were done in service mode and the
cluster was observed on 25 separate nights, starting on 2001 March 12th
and ending on 2001 July 9th.
 
ISAAC, like SofI, is equipped with a Hawaii 1024x1024 HgCdTe array,
which, in the S2 imaging mode, results in a pixel scale of 0\farcs1484
per pixel and a field of view of 2\farcm5.  The cluster was imaged in
$J_{\mathrm s}$ and $K_{\mathrm s}$. The $J_{\mathrm s}$ filter is
slightly narrower and redder than typical $J$ filters and the blue edge
of the filter is defined by the filter and not the atmosphere. This
results in more stable photometry at the expense of introducing a
small colour term.
 
We did not center the cluster within the ISAAC field of view. Instead
we used a series of pointings which put the cluster into the four
corners of the array. The resulting union of images covers a
region which is slightly smaller than the region covered by the SofI
data, but is significantly deeper and has significantly better image
quality.
 
Individual integrations lasted 15 seconds in $K_{\mathrm s}$ and 30
seconds in $J_{\mathrm s}$, and four of these were averaged to form a
single image. Between images, the telescope was offset by 10\arcsec\
to 30\arcsec\ in a semi-random manner, and typically 30 to 40 images
were taken in this way in a single observing block. In total,
approximately 6 hours in both $K_{\mathrm s}$ and $J_{\mathrm s}$ were
spent at each pointing.  Since the cluster center was visible in all
pointings, approximately 24 hours is spent on the cluster in both
$J_{\mathrm s}$ and $K_{\mathrm s}$.

In addition to these deep images, two additional regions that flank
the eastern and western edges of the mosaic were observed so that
potential cluster galaxies could be identified for later spectroscopic
follow-up (Demarco et al. in preparation). The total exposure time
in each of the these flanking fields is 18 minutes, so the depth in
these images considerably shallower that the depth obtained in the
center of the mosaic. The details are given in Table
\ref{tab:ISAAC_observations}.
 
\begin{table*}
\caption[ISAAC data]{A summary of the observations taken with
ISAAC. The exposure times and detection limits of the mosaic refer to
the central part of the mosaic, where the exposures are deepest. In
other areas, the exposure time will vary from one quarter to one half
of this depending on the overlap. The detection limit is defined as
the $5\sigma$ detection threshold over an 0\farcs9 diameter aperture.}
\label{tab:ISAAC_observations}
\begin{tabular}{llrcc}
\hline\hline

Filter & Region      & Exposure & Image Quality & Detection Limit\\
       &             & (seconds)& (\arcsec ) &   (Vega magnitudes)\\
\hline
$K_{\mathrm s}$     & Mosaic   & 81990    & 0.45          & 24.1    \\
$J_{\mathrm s}$     & Mosaic   & 86640    & 0.45          & 25.6    \\
$K_{\mathrm s}$     & Eastern  &  1080    & 0.40          & 21.5    \\
$J_{\mathrm s}$     & Eastern  &  1080    & 0.43          & 23.3    \\
$K_{\mathrm s}$     & Western  &  1080    & 0.32          & 21.5    \\
$J_{\mathrm s}$     & Western  &  1080    & 0.40          & 23.3    \\

\hline   
\end{tabular}
\end{table*}   

The data were reduced in a similar way to that used for the SofI data,
but with following additional steps.
 
\begin{itemize}

\item The difference in the relative level between odd and even
columns was removed. The relative difference is a function of the
average count level and it evolves with time.
 
\item An electronic ghost, which is most easily seen when there are bright
stars in the field of view, was removed.
 
\item Sky subtraction is done with object masking. We used the XDIMSUM package
for this step.
 
\item Images were corrected for field distortion, which can amount to
several pixels at the edges of the ISAAC field of view.
 
\item Individual images were scaled to a common ZP before being combined. 

\end{itemize}

The atmospheric conditions at the time the data were taken were very
good. All but two of the nights were photometric and the image quality
on the raw images varied between 0\farcs25 to 0\farcs8, with a median
around 0\farcs45.  Zero points were derived from the observations of
photometric standards from the catalogue of Persson et al. (1998). The
ISAAC $J_{\mathrm s}-K_{\mathrm s}$ colours were transferred to the
LCO (Persson) system using the transformation

$$ (J-K_{\mathrm s})_{\rm LCO} = 1.028 * (J_{\mathrm s}-K_{\mathrm s})_{\rm ISAAC} - 0.011. $$

As with the SofI data, the colours and magnitudes were then
transferred to the 2MASS system (Carpenter \cite{Carpenter01}) and
only non-stellar sources are corrected for galactic extinction.

The ISAAC data were taken over a large number of nights and,
consequently, the image quality over the entire mosaic is slightly
variable. The dispersion in the image quality as measured from bright
stars over the entire mosaic is about 8\% in $K_{\mathrm s}$ and 5\%
in $J_{\mathrm s}$. The median ellipticity is 0.05 and 0.03 for the
$K_{\mathrm s}$ and $J_{\mathrm s}$ images respectively. In the
central part of the mosaic, the uniformity of the PSF is considerably
better, since the central part of the mosaic is common to all images.

We used the SExtractor software (Bertin \& Arnouts \cite{Bertin96}) to
detect objects, to do the photometry and to classify sources as either
point like or extended. The $J-K_{\mathrm s}$ colour for objects in
the central part of the mosaic are derived from the flux within fixed
apertures of 6 pixels (0\farcs89 diameter). Since the image quality in
the central part of the $J_{\mathrm s}$ and $K_{\mathrm s}$ ISAAC
mosaics are very similar, there are no aperture corrections to the
$J-K_{\mathrm s}$ colours. The photometric errors were calculated
independently, since the error in a fixed aperture is larger than the
error that one would derive by multiplying the noise in a single pixel
by the square root of the number of pixels in the aperture (Labb\'{e}
et al. \cite{Labbe03}). {The total magnitude was estimated using the
``BEST'' magnitude in SExtractor (Bertin \& Arnouts
\cite{Bertin96}).


The independently calibrated SofI images are used to check the
photometric accuracy of the ISAAC data. The difference in the absolute
calibrations are less than 0.02 magnitudes and the scatter in the
$J-K_{\mathrm s}$ colour of relatively bright objects ($K_{\mathrm s}
< 19$) is 0.02 magnitudes over a region that is within 30\arcsec\ of
the cluster center. The scatter increases to 0.03 magnitudes if all
objects within the region covered in Fig. \ref{fig:ISAAC_bw} are
included.  The scatter increases with increasing area because image
quality becomes more variable as one moves away from the cluster
center and all colour measurements are based on aperture
magnitudes. Unless otherwise stated, we restrict the measurement of
cluster properties in the analysis that follows to those galaxies that
lie within 20\arcsec\ of the cluster center.  This region has
been selected as a compromise between maximising the number of cluster
galaxies with respect to the number of field galaxies and obtaining a
sufficient number of cluster galaxies for analysis.  

We also cross checked the photometry of the brightest cluster
galaxies. Although there is significant dispersion in the total
magnitude, about 0.15 magnitudes, the dispersion in the $J-K_{\mathrm
s}$ colour is less than 0.02 magnitudes and the mean difference is
less than 0.01 magnitudes. The comparison between the independently
calibrated SofI and ISAAC data and the comparison between objects in
the SofI data and in the 2MASS catalogue suggest that the systematic
error in our $J-K_{\mathrm s}$ colours is no larger than 0.03
magnitudes.
 
The central 75\arcsec\ of the ISAAC $K_{\mathrm s}$ band image is
shown in Fig. \ref{fig:ISAAC_bw}. This images can be compared to the
image taken with SofI (Fig. \ref{fig:image_SOFI}), and it
demonstrates the increased depth and spatial resolution of the ISAAC
data.

\begin{figure*}
\centering
\setlength{\unitlength}{0.33203mm}
\begin{picture}(512,512)
\includegraphics[width=17cm]{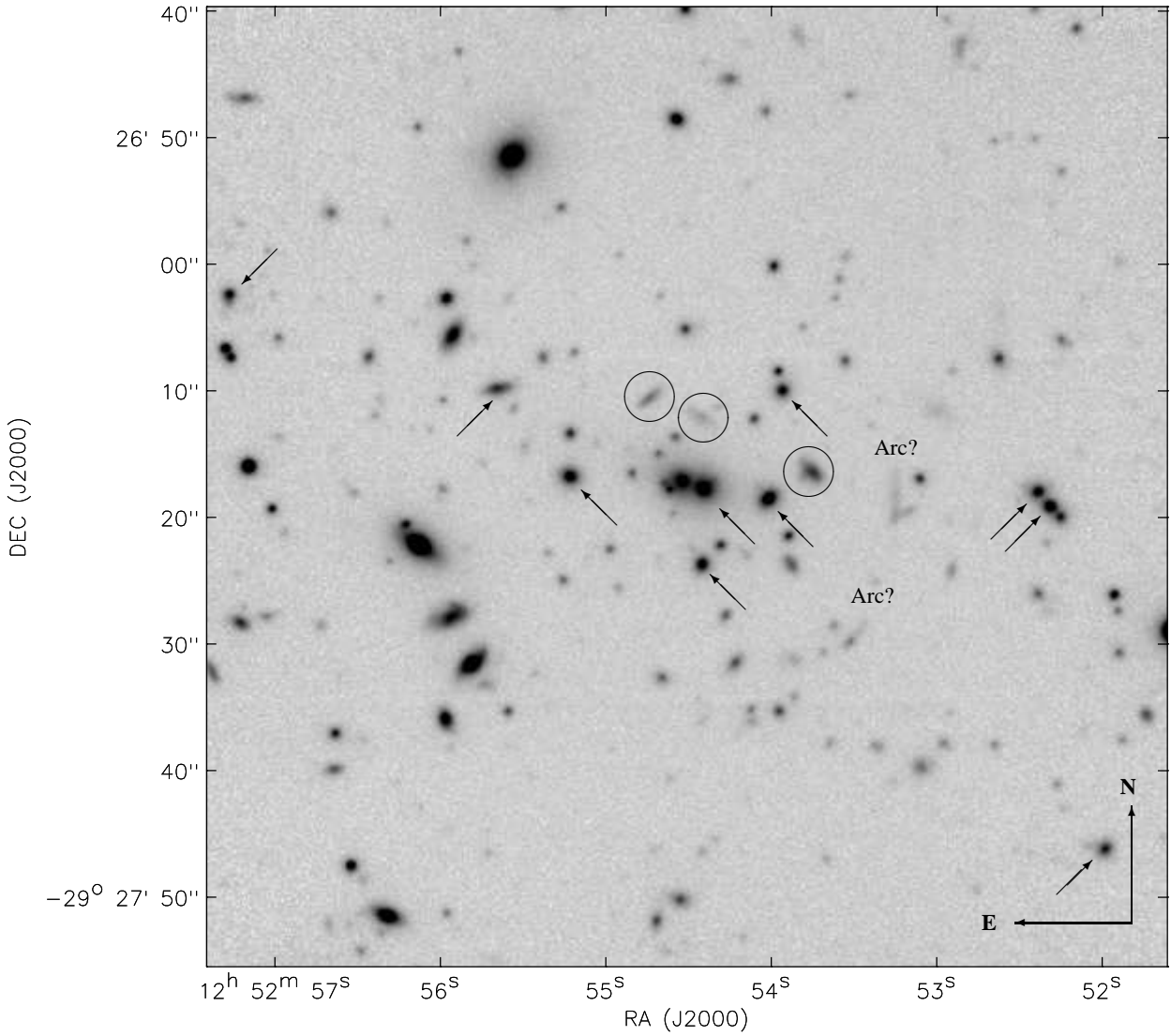}
\put(-170,278){\vector(-1,1){15}}
\put(-260,240){\vector(-1,1){15}}
\put(-201,232){\vector(-1,1){15}}
\put(-205,204){\vector(-1,1){15}}
\put(-176,231){\vector(-1,1){15}}
\put(-72,82){\vector(1,1){15}}
\put(-100,234){\vector(1,1){15}}
\put(-94,229){\vector(1,1){15}}
\put(-328,278){\vector(1,1){15}}
\put(-405,358){\vector(-1,-1){15}}
\put(-223,286){\circle{20.0}}
\put(-246,295){\circle{20.0}}
\put(-178,263){\circle{20.0}}
\put(-40,70){\vector(0,1){50}}
\put(-40,70){\vector(-1,0){50}}
\put(-45,125){\bf N}
\put(-104,67){\bf E}
\put(-150,270){Arc?}
\put(-160,207){Arc?}
\end{picture}
\caption{The central part of the ISAAC $K_{\mathrm s}$-band image. The
image is 75\arcsec\ on a side, which, in the adopted cosmology,
corresponds to 0.67 Mpc at $z=1.237$.  The two galaxies near the
center of the cluster are 1\farcs8 apart. Spectrally confirmed
cluster members and field galaxies (Demarco et al. in preparation) are marked
with arrows and circles respectively. Two arc-like features are also
marked.}
\label{fig:ISAAC_bw}
\end{figure*}      

We mark the spectrally confirmed cluster members (Demarco et
al. in preparation) that lie fully within Fig. \ref{fig:ISAAC_bw}
with arrows. With the exception of one cluster galaxy, which has the
morphology of an edge on spiral and [OII] emission, the
morphologies of the spectrally confirmed cluster members are
consistent with the morphologies of early-type galaxies. None of these
galaxies show detectable [OII] emission (Demarco et
al. in preparation). For comparison, three spectrally confirmed
field galaxies that lie with 20\arcsec\ of the cluster center are
circled. The colours and morphologies of these three galaxies are quite
different to the colours and morphologies of the spectrally confirmed
cluster members. One appears to be slightly disturbed edge-on spiral
and the other two have highly irregular morphologies.

There appears to be two relatively red arc-like features about
20\arcsec\ from the cluster center. Their shape and distance from the
cluster are suggestive of giant gravitational arcs. However, higher
resolution imaging and spectroscopy would be needed before conclusions
could be drawn.

\section{Results}

\subsection{The colour magnitude diagram}

The ISAAC C-M diagram of objects within 20\arcsec\ of the cluster
center is shown in Fig. \ref{fig:CM_ISAAC}. Objects are plotted as
solid symbols if they were classified as extended or as a star if they
were classified as a point source. Objects fainter than $K_{\mathrm s}
< 21$ are not classified and are plotted as solid symbols. Unlike the
C-M diagram that was derived from the SofI data
(Fig. \ref{fig:CM_SOFI}) and for reasons of clarity, objects
outside this radius are not plotted. If they were, one would see a
similar peak in the colours of stellar-like objects around
$J-K_{\mathrm s} \sim 0.8$.

Spectrally confirmed cluster members and field galaxies are marked with
circles and crosses respectively. The field galaxies are also marked
in Fig. \ref{fig:ISAAC_bw}, and all of them are morphologically
distinct from the spectrally confirmed cluster members.

Galaxies that are within 20\arcsec\ of the cluster center lie on a
well defined C-M relation. The line in Fig. \ref{fig:CM_ISAAC} is a
fit to the galaxies within the blue rectangle (i.e. galaxies with
$K_{\mathrm s} < 21$, $J-K_{\mathrm s} > 1.5$ and $J-K_{\mathrm s} <
2.1$) and within 20\arcsec\ of the cluster center. Even thought the
fit has been done for galaxies brighter than $K_{\mathrm s}=21$, the
upper envelope in the colour of galaxies that are as faint as
$K_{\mathrm s}=24$ is defined by this relation. We also indicate, with
boxes in Fig. \ref{fig:CM_ISAAC}, the location of four spectrally
confirmed cluster galaxies that are more than 20\arcsec\ form the
cluster center but within the area bounded by Fig.
\ref{fig:ISAAC_bw}. These cluster galaxies also lie on the C-M relation.

\begin{figure*}
\centering
\includegraphics[width=17cm]{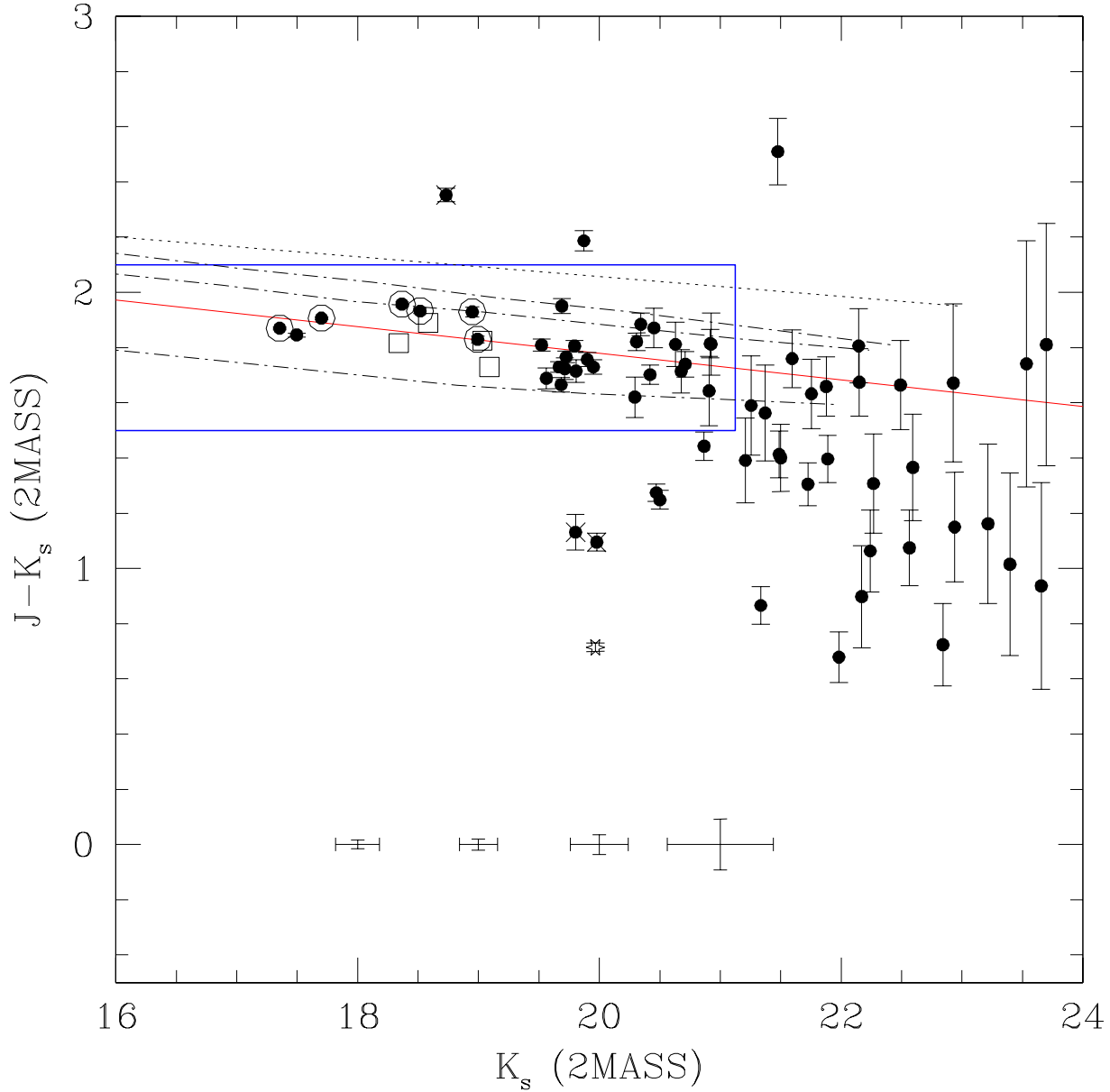}
\caption{The C-M diagram of objects within 20\arcsec\ of the cluster
center. The diagram is generated from the ISAAC data. The solid
circles represent objects that have been classified as extended and
the stellar symbol represents the object that was classified as a
star.  Spectrally confirmed cluster members are circled and spectrally
confirmed field galaxies are marked with an ``X''. The solid red line
is a fit to the C-M relation of galaxies within the blue rectangle
(see text) and the dotted line is where the E/S0 sequence of the Coma
cluster would lie if it were placed at z=1.24 (Rosati et
al. \cite{Rosati99}). The dot-dashed lines are monolithic collapse
models (Kodama \& Arimoto \cite{Kodama97}). From top to bottom they
represent formation redshifts of {\bf $z=5,3\, {\rm and }\, 2$}. The error
bars that are located near the bottom of the plot indicate the size of
the colour and magnitude errors. Also plotted, as open squares, are
the spectrally confirmed cluster members that are more than 20\arcsec\
from the cluster center but within the field of view of Fig.
\ref{fig:ISAAC_bw}.}
\label{fig:CM_ISAAC}
\end{figure*}

Recently, Labb\'{e} et al (\cite{Labbe03}) measured the galaxy number
counts to very faint NIR magnitudes. We have used their results to
estimate that, to $K_{\mathrm s}=21$, six field galaxies are expected
within 20\arcsec\ of the cluster center, which is considerably fewer
than the 36 galaxies we do observe.  We do not separate field and
cluster galaxies when fitting the C-M relation, since the three
spectrally confirmed field galaxies are already excluded by our colour
cuts, i.e. they lie outside the blue rectangle in Fig.
\ref{fig:CM_ISAAC}, and the remaining small number of unidentified
field galaxies are unlikely to bias our fit.

The fit to the C-M relation was determined by adjusting the amount of
intrinsic scatter that had to be added to the data until the reduced
$\chi^2$ was one. A floor to the error in the colour, as determined by
comparing the colours of the brightest galaxies in the ISAAC and SofI
data, was set to 0.02 magnitudes.  Only those objects within the blue
rectangle in Fig. \ref{fig:CM_ISAAC} were used in the fit. Separate
fits were done for regions of different radii centered on the
cluster. In all cases, the dominant source of scatter is the intrinsic
scatter. For completeness, we also report the fit to the slope of the
C-M relation without adding additional scatter. The fits, the observed
scatter and the inferred intrinsic scatter are reported in Table
\ref{tab:CM_fit}.

\begin{table*}
\caption[CM fits]{The fit to the C-M relation. Only
objects within the blue rectangle of Fig. \ref{fig:CM_ISAAC} are used
in the fit.}
\label{tab:CM_fit}
\begin{tabular}{lllllll}
\hline   
\hline   
Radius & Number of & Slope$^1$& Slope$^2$            &Observed &Intrinsic & 90\% confidence region\\
(\arcsec )&Objects &          &                      & Scatter &Scatter   & for intrinsic scatter \\
\hline
10    & 14        & -0.046    & -0.058 $\pm$ 0.017 &   0.066 &   0.060  & 0.039 to 0.091  \\
15    & 19        & -0.052    & -0.055 $\pm$ 0.015 &   0.068 &   0.058  & 0.041 to 0.086  \\
20    & 29        & -0.046    & -0.048 $\pm$ 0.015 &   0.082 &   0.070  & 0.052 to 0.093  \\
\hline
\end{tabular}
\newline
Notes:\\
$^1$ No added scatter\\
$^2$ With additional scatter\\
\end{table*}   

In measuring the intrinsic scatter, it is imperative that measurement
errors are accurately estimated. We checked the accuracy of our
measurement errors by adding artificial galaxies directly to the
reduced data and by processing the artificial and real data in an
identical manner. The errors estimated in this way are in excellent
agreement with the errors that are estimated from the image noise.
Representative error bars are plotted in lower part of Fig.
\ref{fig:CM_ISAAC} for galaxies with apparent magnitudes of
$K_{\mathrm s}=$ 18, 19, 20 and 21.

We also used this technique to check for biases in the
photometry. There is a tendency for the SExtractor BEST magnitude to
miss an increasing fraction of the flux as objects become fainter.
However, the error is less than 0.2 magnitudes over the magnitude
range in which the fit to the C-M relation is done.  Aperture
magnitudes and the colours that are derived from them are unbiased.

The slope of the C-M relation is -0.05 magnitudes per magnitude, which
is quite similar to the slope measured in clusters with redshifts up
to $z \sim 0.9$ (SED). If the fit is done using aperture magnitudes
instead of the total magnitudes, the slope steepens slightly, although
the change is within the measurement error. The scatter about the
relation is unchanged.  If the fit was repeated just for the
spectrally confirmed cluster galaxies that are marked in figure
\ref{fig:ISAAC_bw}, the fitted slope and the measured scatter are also
similar.



Recent observations of clusters above $z\sim 1$ have pointed to a
possible flattening in the slope (van Dokkum et
al. \cite{vanDokkum01b}; Stanford et al. \cite{Stanford02}). We see no
evidence for a flattening in the slope of the C-M relation in
\object{RDCS J1252.9-2927}. The dotted line in Fig.
\ref{fig:CM_ISAAC} is how the C-M relation of the Coma cluster would
appear if it were moved to the redshift of \object{RDCS J1252.9-2927}
(Rosati et al. \cite{Rosati99}). Within the measurement uncertainties
the slope for \object{RDCS J1252.9-2927} is the same, but the
$J-K_{\mathrm s}$ colours of galaxies in \object{RDCS J1252.9-2927}
are on average 0.25 magnitudes bluer. This result is consistent with
the trend seen in SED.

We estimate the intrinsic scatter to be 0.06 magnitudes with a 90\%
confidence interval from 0.04 to 0.09 magnitudes. This is similar to
the scatter seen in clusters with redshifts up to $z\sim 0.9$
(SED). Even, if we have grossly overestimated the measurement
errors, which is unlikely, the intrinsic scatter cannot be much larger
than the measured scatter (0.06-0.08 magnitudes).

Within 20\arcsec\ (0.16 Mpc) of the cluster center, 90\% of galaxies
brighter than $K_{\mathrm s}=21$, which is approximately 2.5
magnitudes fainter than the apparent magnitude of an $L^{\star}$
galaxy in this cluster (Toft et al. in preparation) and 3.5 magnitudes
fainter than the brightest cluster members, lie on the C-M
relation. Three of the seven galaxies that do not lie on the sequence
were found to be non-cluster members from follow-up spectroscopy. Of
the other four, one is redder than the C-M relation and the other
three are fainter than $K_{\mathrm s}=20$. There is no progenitor bias
in the central regions of this cluster.

The average colour of a $L^{\star}$ galaxy in \object{RDCS
J1252.9-2927} is $J-K_{\mathrm s} \sim 1.85$. This is similar to the
colours measured for the early-type galaxies in RX J0848.9+4452 at
$z=1.26$ (Rosati et al. \cite{Rosati99}), which have $J-K \sim 1.85$
on the UKIRT system. However, the transformation between the UKIRT and
2MASS systems is uncertain, so it is not clear how comparable the
average colour of galaxies in these two clusters are. Using the
transformation equations in Hawarden et al. (\cite{Hawarden01}), the
colours agree very well, but using the transformation equations of
Carpenter (\cite{Carpenter01}), one would find that the early-type
galaxies RX J0848.9+4452 are 0.1 magnitudes redder.

\subsection{The epoch of galaxy formation}

We have used the monolithic collapse models of Kodama and Arimoto
(\cite{Kodama97}) to illustrate how the C-M relation changes with
formation redshift.  In these models, star formation occurs during gas
infall and terminates when the energy from supernovae ejecta exceeds the
binding energy of the gas. Larger galaxies are able to hold onto their
gas longer, and, consequently, star formation occurs over a more
extended period of time and they have higher mean metallicities. Three
models (with $z_{\mathrm f}$ = 2,3 and 5, where $z_{\mathrm f}$ is the
redshift of formation) are plotted in Fig. \ref{fig:CM_ISAAC} as the
dot-dashed lines. They reproduce the measured slope very well and
would suggest that the bulk of the stars in this cluster formed
between $z=2$ and $z=3$.

We have also used the simple stellar population (SSP) synthesis models
of Bruzual and Charlot (\cite{Bruzual03}) to estimate the mean age and
the age spread of galaxies in the center of \object{RDCS
J1252.9-2927}. We use $J-K_{\mathrm s}=1.85$, which is the average
colour of a $L^{\star}$ galaxy in the cluster, and a spread in colours
from $J-K_{\mathrm s}=1.79$ to $J-K_{\mathrm s}=1.91$ to estimate the
mean age and the age spread. If we assume that the spread in colours
is entirely due to age difference, then, for instantaneous
single-burst models with solar metallicity, a mean age of
approximately 2.7 Gyrs and an age spread from 2.2 to 3.2 Gyrs are
derived. These correspond to a mean formation redshift of $z_{\mathrm
f}=2.8$ and a spread in formation redshifts from $z_{\mathrm f}=2.4$
to $3.6$. The upper limits are highly uncertain. An increase of $0.01$
magnitudes in $J-K_{\mathrm s}$, which is smaller than our estimate of
the systematic error, will move the upper limit beyond $z_{\mathrm
f}=5$. The model colours in these simple models do not get much redder
than $J-K_{\mathrm s}=1.93$ for solar metallicities.

If these models are allowed to evolve passively to $z=0.5$, the
corresponding scatter in $J-K_{\mathrm s}$ evolves to less than 0.01
magnitudes, which is considerably less than the scatter measured in
SED. However, we caution that there are differences in the way C-M
relations in this paper and Stanford et al. have been built. The C-M
relation in this paper is created from the central 0.33 Mpc of the
cluster, whereas those of SED are created from larger areas.  With
additional optical and spectroscopic data on \object{RDCS
J1252.9-2927}, one will be able to extend the area over which the C-M
relation is derived so that a more robust comparison can be made.


Since the observed scatter in $J-K_{\mathrm s}$ does not change from
$z=1.237$ to the present day, and since the scatter in SSP and
monolithic collapse models should decrease as time goes on, it seems
unlikely that the observed scatter in \object{RDCS J1252.9-2927} is
solely due to age differences. Alternative reasons for the scatter are
metallicity variations at constant luminosity and/or dissipationless
merging with little subsequent star formation. Such mergers may have
already been identified in both cluster and field environments
(vanDokkum et al. {\cite{vanDokkum99}; van Dokkum et
al. \cite{vanDokkum01b}; van Dokkum et al. \cite{vanDokkum03b}).

If most of the scatter is due to metallicity differences, the mean age
of formation does not change, but the range of redshifts over which
galaxies would have formed narrows. If most of the scatter is due to
dissipationless merging, then, in the monolithic collapse models of
Kodama and Arimoto (\cite{Kodama97}), the mean age increases and the
formation redshift is pushed higher, because galaxies will move left
and down in the C-M diagram after the merging has taken place. The
directions that galaxies move depend on the relative masses and
metallicities of the progenitors.

The derived formation epoch is directly related to the assumed
metallicity and the star formation history. If a lower metallicity was
assumed, an earlier formation epoch would be derived. This is the well
known age-metallicity degeneracy, and the NIR imaging data that are
presented here cannot break the degeneracy. An advance in this area
will require a comparison between deep spectroscopic data and the most
recent stellar population synthesis models (Bruzual \& Charlot
\cite{Bruzual03}). However, in the SSP models that have been used
here, the galaxies in \object{RDCS J1252.9-2927} cannot be much metal
poorer than solar, otherwise the formation epoch is pushed beyond the
big bang. Additionally, the metallicity of the intra-cluster medium in
this cluster is $\sim 0.4$ times solar (Rosati et al. \cite{Rosati03a}).


\section{Conclusions}

We have obtained very deep, $J-$ and $K_{\mathrm s}$-band images of
the X-ray luminous galaxy cluster \object{RDCS J1252.9-2927} at
$z=1.237$ with ISAAC on the ESO VLT and with SofI on the ESO NTT. The
data enable us to construct a $J-K_{\mathrm s}$ versus $K_{\mathrm s}$
C-M diagram to $K_{\mathrm s}=24$, which is five magnitudes below
$L^{\star}$ (Toft et al. in preparation).

Galaxies within 20\arcsec\ of the cluster center define a tight C-M
relation. The slope of the relation is -0.05 magnitudes per magnitude
and is similar to the slope measured in clusters at lower redshifts
(SED). This strengthens the hypothesis that the slope in the C-M
relation is due to metallicity and not age. We see no evidence for a
flattening in the slope as predicted in hierarchical models and
tentatively observed in clusters at $z \sim 1$ (van Dokkum et
al. \cite{vanDokkum01b}; Stanford et al. \cite{Stanford02}).

More than 90\% of the galaxies within 20\arcsec\ of the cluster center
and brighter than $K_{\mathrm s}=21$ lie on the C-M relation. There is
no progenitor bias in the centre of this cluster.

The intrinsic scatter in the $J-K_{\mathrm s}$ colour of galaxies
about the C-M relation in \object{RDCS J1252.9-2927} is 0.06
magnitudes and is similar to the scatter measured in clusters from
$z=0$ to $z\sim 0.9$ (SED). Hence, the scatter has not evolved from
$z=1.24$ to the present day. This weakens the hypothesis that the
scatter in the C-M relation is solely due to age. Dissipationless
merging and metallicity variations at constant luminosity could also
contribute to the scatter. We also see no evidence for increased
scatter in the colours of galaxies at the bright end of the
C-M relation.

Our results can be compared to those derived from high-resolution
optical images of \object{RDCS J1252.9-2927} that were taken with the
Advanced Camera for Surveys on the Hubble Space Telescope in the
F775W and F850LP filters. BFP find a tight C-M relation in
the $i_{\mathrm 775}-z_{\mathrm 850}$ versus $z_{\mathrm 850}$ C-M
diagram, and they show that neither the slope of this C-M relation nor
the scatter about it have evolved from $z=0$ to $z=1.24$. This concurs
with the findings of this paper. 

Using instantaneous, single-burst, solar-metallicity models, the
average age of the bulk of the stars in the center of the cluster is
2.7 Gyrs. This corresponds to a formation redshift of $z_{\mathrm
f}=2.8$. If the scatter about the CM relation is due to age, most of the
galaxies in the center of this cluster were formed between $z=2.4$ and
$z=3.6$.

\begin{acknowledgements}

This work would not have been possible without the dedicated efforts
of ESO staff, both in Chile and Europe. We are grateful to
Dr. Tadayuki Kodama for providing us with a copy of his models. We'd
also like to thank Drs. Gustavo Bruzual and Stephane Charlot for
providing us with a copy of their stellar evolution models prior to
publication. Part of this work was performed under the auspices of
the U.S.  Department of Energy by University of California, Lawrence
Livermore National Laboratory under contract No. W-7405-Eng-48.

\end{acknowledgements}

\end{document}